\documentclass{icrc}

\usepackage{times}
\usepackage{graphicx}

\begin{document}

\title{New developments in the GALPROP CR propagation model }
\author[1]{A. W. Strong}
\affil[1]{Max-Planck-Institut f\"ur extraterrestrische Physik, 
Postfach 1312, 85741 Garching, Germany}
\author[2,3]{I. V. Moskalenko}
\affil[2]{NRC-NASA/Goddard Space Flight Center, Code 660, Greenbelt, MD 20771, U.S.A.}
\affil[3]{Institute of Nuclear
   Physics, M.\ V.\ Lomonosov Moscow State University, 119 899 Moscow, Russia}

\correspondence{aws@mpe.mpg.de}

\firstpage{1}
\pubyear{2001}

% \titleheight{11cm} % uncomment and adjust in case your title block
                     % does not fit into the default and minimum 7.5 cm

\maketitle

\def\gray{$\gamma$-ray\ }
\def\grays{$\gamma$-rays\ }

\begin{abstract}

The GALPROP cosmic-ray (CR) propagation model has been extended to
three dimensions including the effects of stochastic SNR sources,
a comprehensive cross-section data\-base, and nuclear reaction networks. 
A brief description of the new code is given and some illustrative
results for the distribution of CR protons presented.  Results for
electrons and \grays are given in an accompanying paper.
\end{abstract}

\section{Introduction}
 
We have previously described a numerical model for the \linebreak[4]Galaxy
encompassing primary and secondary cosmic rays, \grays and synchrotron
radiation in a common framework (Strong et al.\ 2000 and references
therein). Up to recently our GALPROP code handled 2 spatial dimensions,
$(R,z)$, together with particle momentum $p$. This was used as the
basis for studies of CR reacceleration, the size of the
halo, positrons, antiprotons, dark matter and the interpretation of
diffuse continuum $\gamma$-rays. 

Some aspects of the problem cannot be
addressed in such a cylindrically symmetric model: for example the
stochastic nature of the cosmic-ray sources in space and time, which
is important for high-energy electrons with short cooling times, and
local inhomogeneities in the gas density which can affect radioactive
secondary/prima\-ry ratios.  

In common with most other models it has
been previously assumed that the CR source function can be taken as
smooth and time-independent, an approximation justified by the long
residence time ($>10^7$ years) of cosmic-rays in the Galaxy.  However
the inhomogeneities have observable consequences, and their inclusion
is a step towards of the goal of a ``realistic'' propagation model based
on Galactic structure and plausible source properties.  The original
motivation for this extension was to study the high-energy electrons,
since the observation of the $>1$ GeV excess in the EGRET spectrum of
the Galactic emission has been proposed to originate in
inverse-Compton emission from a hard electron spectrum; this
hypothesis can only be reconciled with the local directly-observed
steep electron spectrum if there are large spatial variations which
make the spectrum in our local region unrepresentative of the
large-scale average.
 
Here we briefly describe an extension of the model to 3D, which can
address these issues, and illustrate the results for protons.
The new network and cross-sections is illustrated for B/C in the 2D case.
The effect on electrons and \grays is presented
in an accompanying paper (Strong and Moskalenko, `A 3D time-dependent
model for cosmic rays and $\gamma$-rays', these proceedings: hereinafter
paper II).

\section{Model} 

The GALPROP code, which solves the CR propagation equations on
a grid, has been entirely rewritten in C++ using the experience
gained from the original version and including both 2D and 3D
spatial grid options. The 2D mode
essentially duplicates the original version, with improved
cross-section routines. In 3D $(x,y,z,p)$ the propagation is solved as
before using a Crank-Nicolson scheme. The additional dimension
considerably increases the computer resources required, but a 200 pc
grid cell or finer is still practicable. As in the original version,
the effects of diffusion, convection, diffusive reacceleration,  and
energy losses are included, each with adjustable parameters defined
for a GALPROP run.  
 
\subsection{Cross-sections and reaction network}
Cosmic-ray nuclear reaction networks are
included with a comprehensive new cross-section database; this allows
the models to be tuned on stable and radioactive CR sec\-ond\-ary/primary
ratios, in particular B/C and $^{10}$Be/$^9$Be. 
%%%%%%%%%%%%%%%%%%%%%%%%%%%%%%%%%%%%%%%%%%%%%%%%%
The nuclear reaction network
is built using the Nuclear Data She\-ets. The isotopic cross section database 
consists of more than 2000 points collected from sources published in 1969--1999.
This includes a critical re-evaluation of some data and cross checks.
The isotopic cross sections for B/C were calculated using the authors' fits to major 
beryllium and boron production cross sections C,N,O $\to$ Be,B.
Other cross sections are calculated using the semi-phenomenological
approximations by \citet{webber} (code WNEWTR.FOR of 1993)
and/or Silberberg and Tsao (code of 2000)
renormalized to the data where it exists.

The reaction network is solved starting at the heaviest nuclei (i.e.\ $^{64}$Ni).
The propagation equation is solved, computing all the resulting secondary so\-urce
functions, and then proceeds to the nuclei with $A-1$. 
The procedure is repeated down to $A=1$.
In this way all secondary, tertiary etc.\ reactions are automatically accounted for.
To be completely accurate for all isotopes, e.g.\ for some rare cases of 
$\beta^\pm$-decay, the whole loop is repeated twice.
Our preliminary results for all cosmic ray species $Z\leq28$ are given in
\citet{SM01}.
%%%%%%%%%%%%%%%%%%%%%%%%%%%%%%%%%%%%%%%%%%%%%%%%%

\subsection{ Stochastic sources}
Another major enhancement is the inclusion of
stochastic SNR events as sources of cosmic rays. The SNR are
characterized by the mean time $t_{SNR}$ between events in a 1 kpc$^3$
unit volume, and the  time $t_{CR}$ during which an SNR actively
produces CR.

The propagation is first carried out for a smooth distribution of
sources to obtain the long timescale solution, using the fast
technique described in \citet{SM98}; then the
stochastic sources are started and propagation followed for the last
10$^7$ years or so with  timesteps sufficiently fine (e.g., 10$^3$
years) to resolve the SNR events and all propagation effects (see
below for more details of the method). For high-energy (TeV)
electrons  which lose energy on timescales of 10$^5$ years the effect
is a very inhomogeneous distribution with consequences for diffuse
$\gamma$-rays, as shown in the accompanying paper II.  However also the
protons (and other nuclei) show fluctuations which are strongly
energy-dependent.  The amplitude of the fluctuations of both electrons
and nuclei depends  on the  parameters $t_{SNR}$ and $t_{CR}$.  The
time $t_{SNR}$  is adjusted to be consistent with  estimates of the
SNR rate (e.g., Dragicevich et al.\ 1999); models for shock acceleration
in SNR indicate $10^4  < t_{CR} < 10^5$  yr, the sources switching
off when they move from the  adiabatic to the radiative phase 
\citep{sturner}.
  
An additional advantage of the 3 spatial dimensions is that the gas
distribution (for energy losses, secondary production and $\gamma$-rays) can
be modelled in as much detail as required, based on current HI, CO,
FIR, etc. surveys. Spiral structure and local inhomogeneities can be
included if required. Global parameters such as the CR luminosity of
the Galaxy, and hence the average SNR energy injection into CR, are
also computed since they provide essential constraints (see paper II).

\subsection{Extension to 3D}

Since the  solution in $(x,y,z,p)$ on a fine grid involves large
arrays and many time-steps, the code has been made vectorizable so
that it can benefit from the use of vector machines with a speed gain
typically 100. This is only required for the final ``stochastic SNR''
part of the solution, since the ``smooth'' solution is fast enough on
non-vector machines. In addition we have the option to make use of
symmetry in the spatial dimension where it does not affect the
application of the results (e.g., about $z=0$). This leads to an
addition large gain in speed. Typical parameters are:  grid size 
$\Delta x=\Delta y=\Delta z= 200$ pc, 
Galaxy dimension 60 kpc $\times$ 60 kpc $\times$ 4 kpc,
and     energy range 1 MeV -- 1 TeV on a logarithmic scale with factor 1.2.
\footnote{As usual the software will be made available on the WWW at
http://www.gamma.mpe--garching.mpg.de/$\sim$aws/aws.html}

\section{CR Protons}

For illustration we show results for a model with reacceleration based
on \citet{SMR00}, and  $t_{SNR}$ = 10$^4$ yr
corresponding to a Galactic rate of 3 SN/century.  Fig.\ 1 shows the
distribution of protons in the Galactic plane ($z=0$) for a
representative quadrant, at various energies.  The stochastic SNR
source produce fluctuations, which are a minimum around 1~GeV and
increase at low energies due to energy losses and at high energies
where the storage of particles in the Galaxy is much reduced so that
the effect of sources  manifests itself on the distribution. Note that
the nature of the fluctuations is different at low and high energies.
 
Fig.\ 2 shows a sample of proton spectra at different $(x,y)$ positions;
the fluctuations are evident but much smaller than for electrons
(paper II).                                                                  

\begin{figure*}[!p]
\hskip -18mm
\includegraphics[height=.3\textheight]{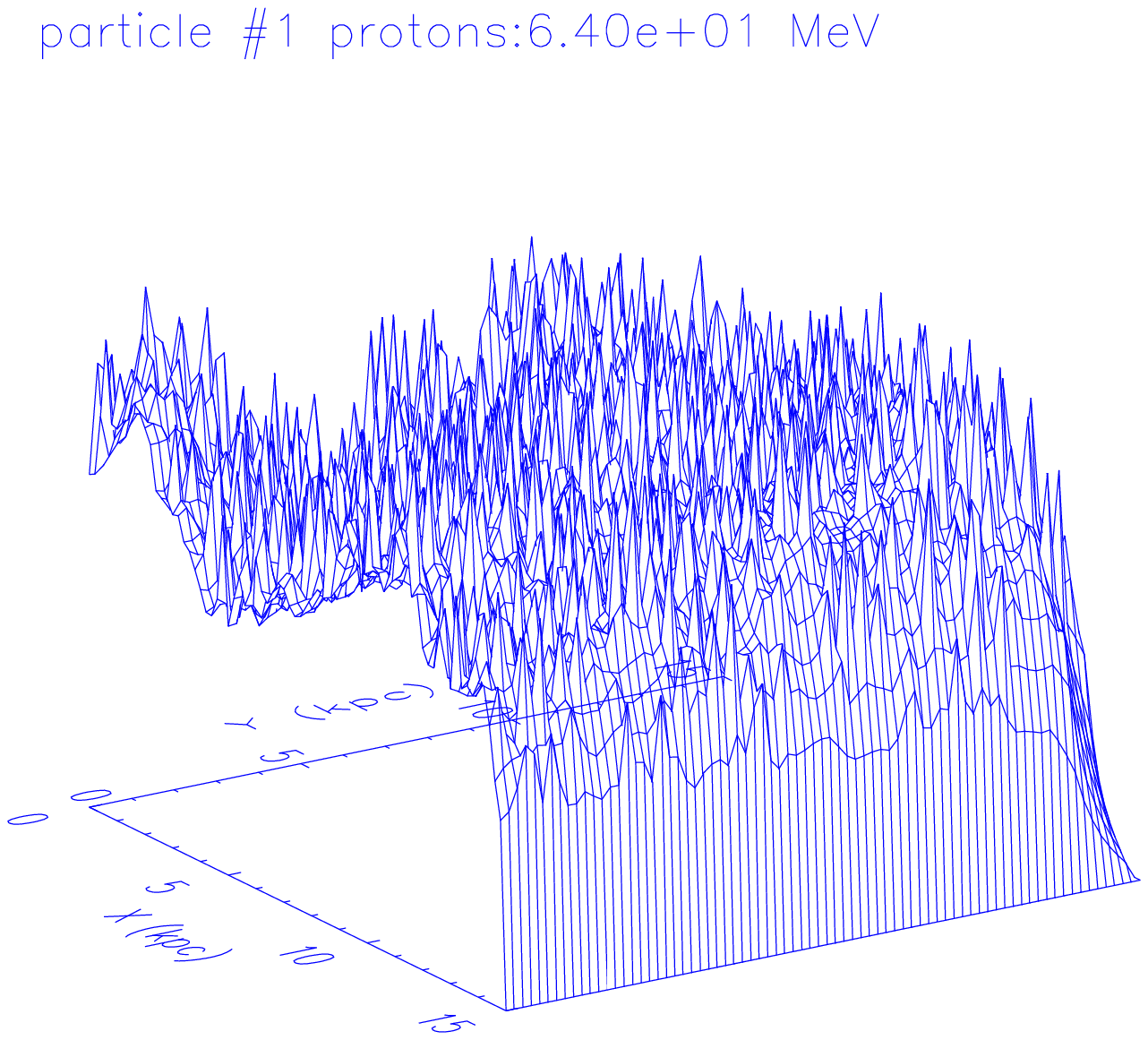}
\includegraphics[height=.3\textheight]{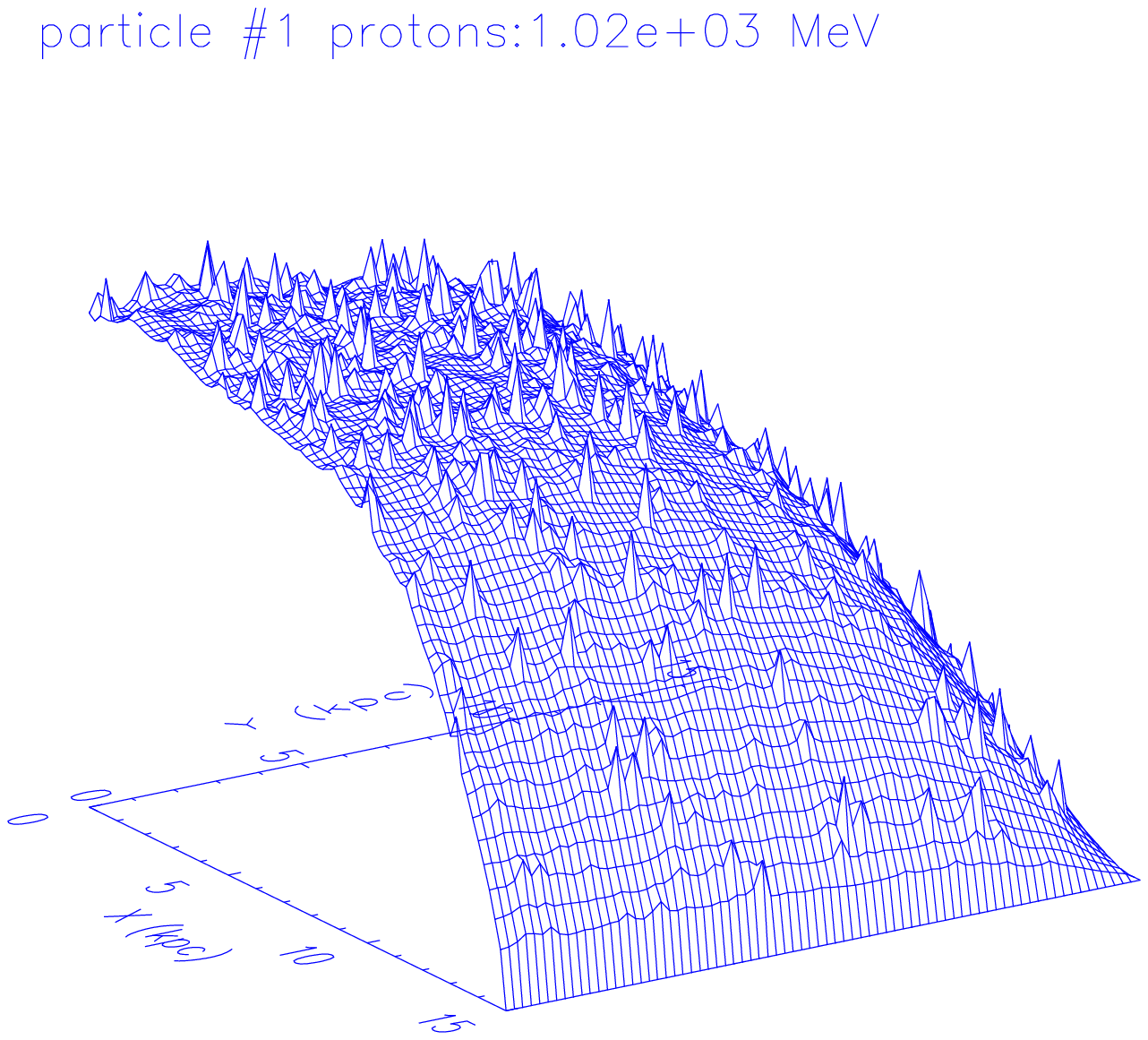}

\hskip -18mm
\includegraphics[height=.3\textheight]{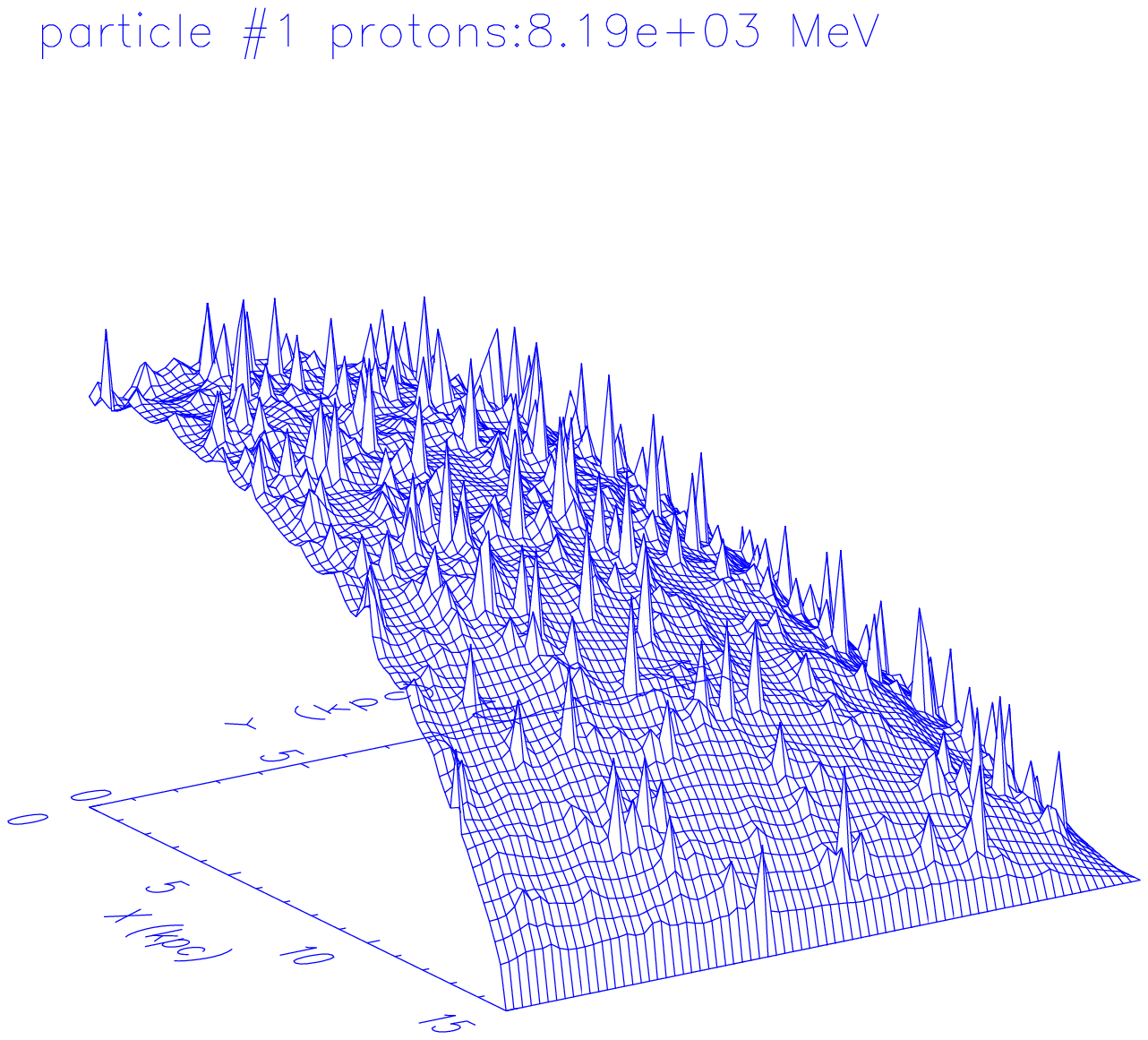}
\includegraphics[height=.3\textheight]{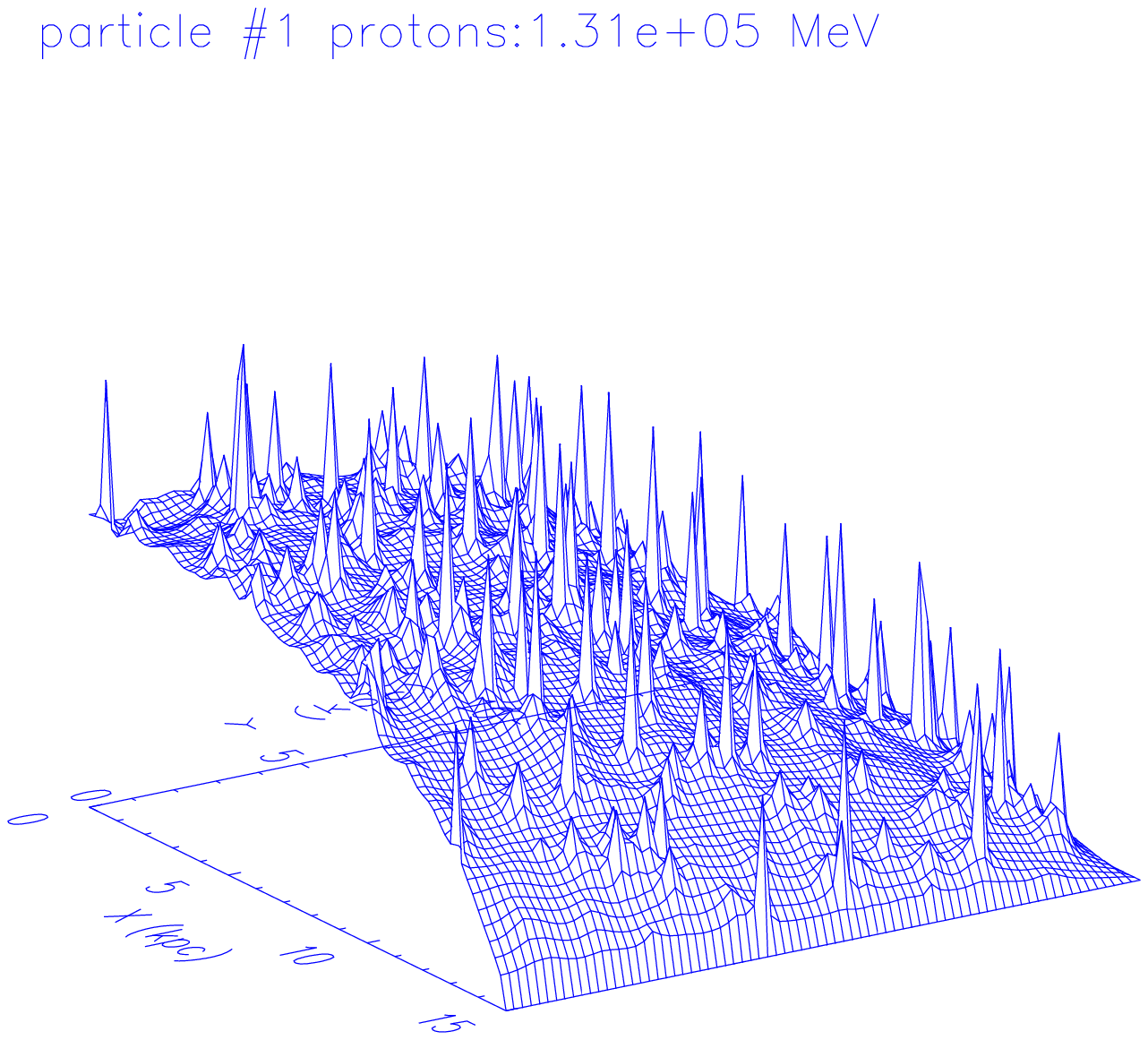}

\centerline{
\includegraphics[height=.3\textheight]{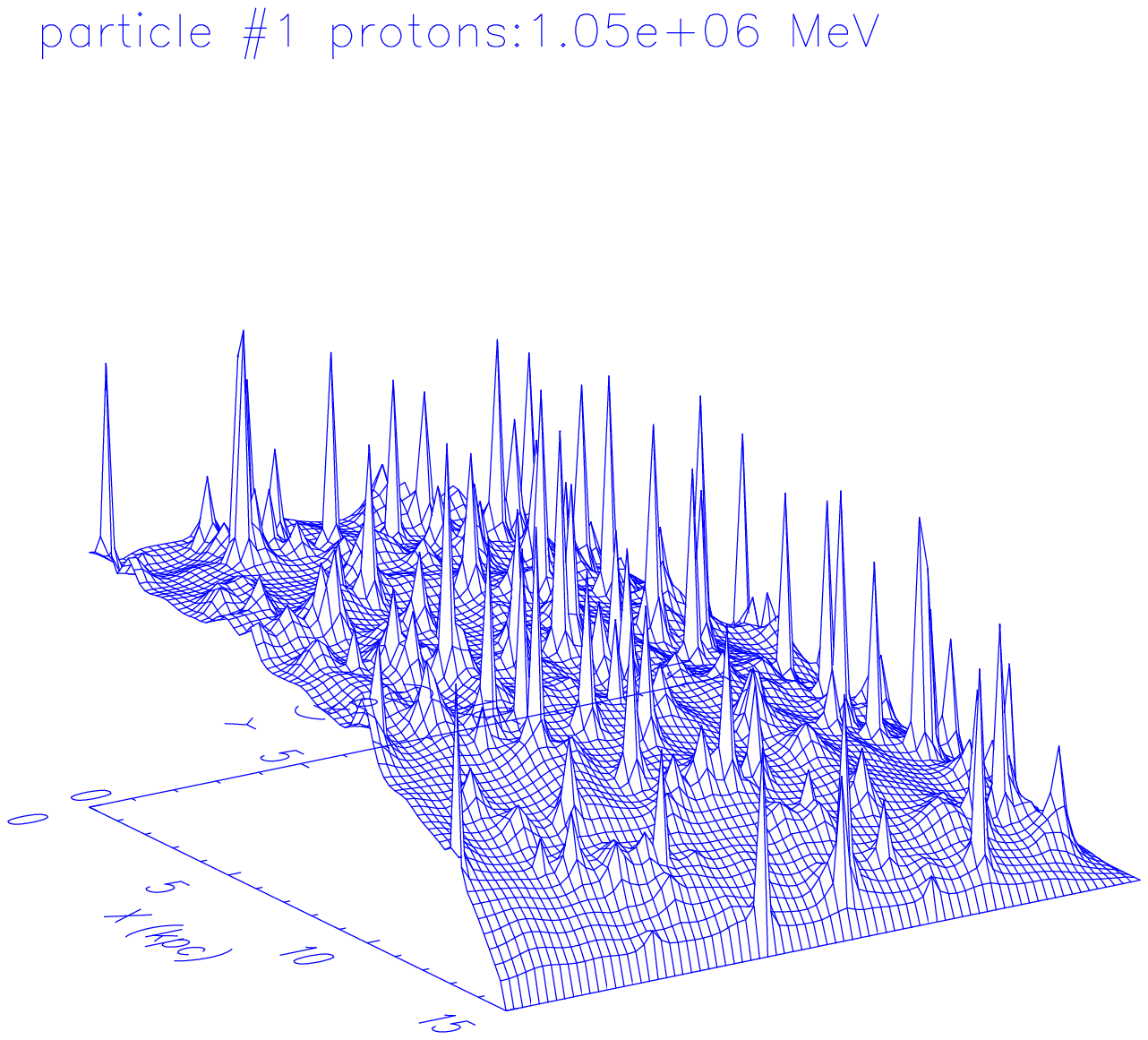}}
\caption{Flux of protons with E= 64 MeV, 1 GeV, 8.2 GeV, 130 GeV, and 1 TeV 
(left to right, top to bottom) at $z=0$,
for $t_{SNR}$ = 10$^4$ yr.   }
\end{figure*}

The fluctuations in the GeV nuclei will have some effect on the
$\pi^0$-decay diffuse \gray emission above 100 MeV, but the
variations will be small compared to the inverse Compton component
since the GeV nuclei variations are much smaller than for the TeV
electrons responsible for GeV \grays via IC (see paper II). Above 100
GeV \gray energies however the effect on the $\pi^0$-decay component
will be larger, and should be detectable for example by the GLAST
mission.

\begin{figure*}[!t]
\includegraphics[height=.37\textheight,width=0.5\textwidth]{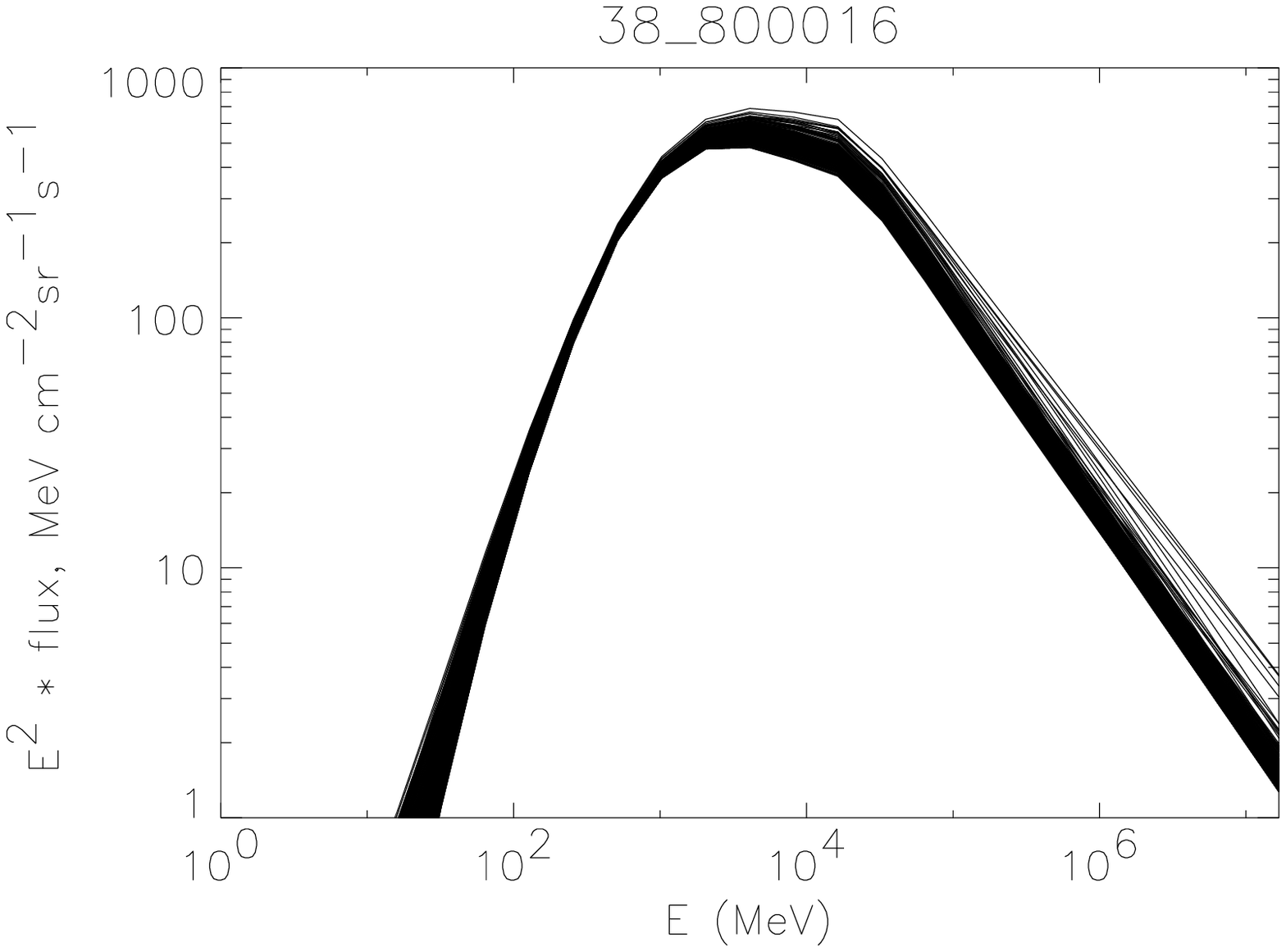}
\includegraphics[width=0.48\textwidth]{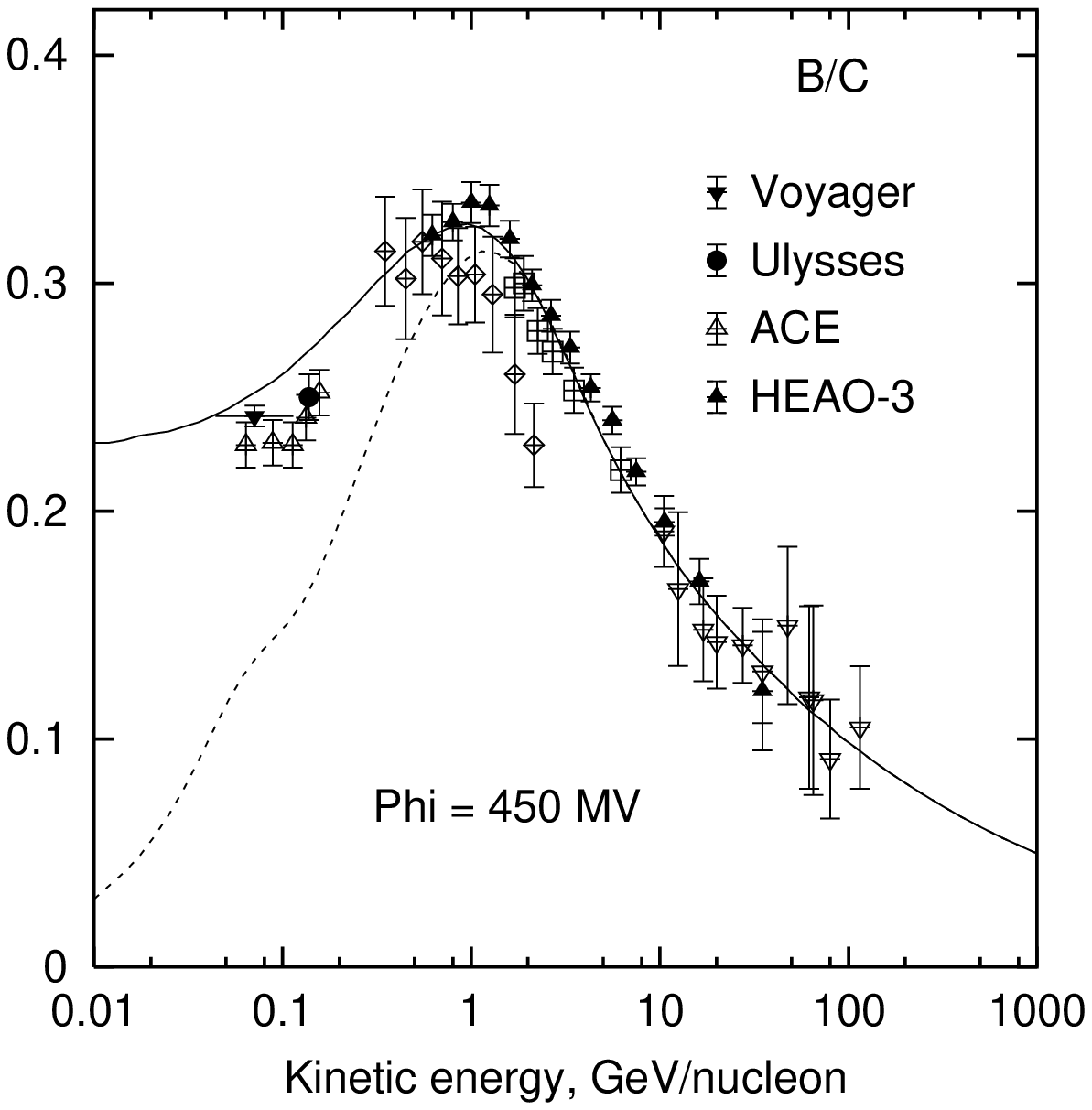}
\begin{minipage}[t]{85mm}
\caption{Sample proton spectra for various $x,y$ at $z=0$, illustrating the variations of 
shape with position. Model as Fig.\ 1.  }
\end{minipage}
\hfill
\begin{minipage}[t]{85mm}
\caption{B/C ratio as calculated for a model with reacceleration using
the new reaction network and cross-sections. Lower curve: interstellar,
upper curve: modulated for 450 MV. Data: see \citet{SM01}.}
\end{minipage}
\end{figure*}

\section{Secondary/primary ratio}

We illustrate the new cross-sections and network with the B/C ratio (Fig.\ 3)
for the case of the same model (with reacceleration) as for electrons.
More details of the application to nuclei can be found in \citet{SM01} and
Moskalenko et al.\
`New calculation of radioactive secondaries in cosmic rays' (these proceedings)
and to antiprotons in Moskalenko et al.\ `Secondary antiprotons in cosmic rays'
(these proceedings).

\section{Conclusion}
This paper is intended only as an illustration of the possibilities opened up
with such a 3D code.  Future work will concentrate on the effect of
inhomogeneities in the gas distribution (e.g., the local bubble) on
radioactive CR nuclei, as well as implications for electrons and
$\gamma$-rays.

\begin{acknowledgements}
IVM acknowledges support from the NRC/NAS Research Associateship Program.
\end{acknowledgements}

\end{document}